\documentclass[5p,times,twocolumn,nopreprintline]{elsarticle}
\usepackage{amsmath,amssymb,amsfonts}
\usepackage{graphicx,subcaption,epsfig}
\usepackage{xcolor}
\usepackage{dcolumn}
\usepackage{multicol}
\usepackage{upquote,multirow,slashed}
\usepackage{appendix}
\usepackage{empheq}
\usepackage{lipsum}
\usepackage{booktabs,dsfont}
\usepackage{dcolumn}
\usepackage{xspace}
\usepackage{balance}

\definecolor{ForestGreen}{rgb}{0.15,0.70,0.15}
\usepackage[colorlinks=true]{hyperref}
\hypersetup{
 linktocpage=true,
 citecolor=ForestGreen,
 linkcolor=blue,
 urlcolor=blue}

\usepackage{fancyhdr}
\fancypagestyle{firstpage}{%
	
	\lhead{}
	\rhead{\small CPTNP-2025-030, MPP-2025-167}
}
\allowdisplaybreaks[1]

\newcommand{\NOdisplay}[1]{ }

\def\MSbar{\overline{\mathrm{MS}}}
\def\SigmaMass{m_{\sigma}}
\def\mMS{\overline{m}}
\def\mOS{m_{\mathrm{os}}}
\def\ZTA{\mathrm{Z}_{\scriptscriptstyle{\mathrm{TA}}}}
\newcommand{\nn}{\nonumber \\}
\newcommand{\api}{\frac{\alpha_s}{\pi}}
\newcommand{\loguos}{L_{\mathrm{os}}}
\newcommand{\logusm}{L_{\sigma}}
\newcommand{\logums}{L_{\mathrm{ms}}}

\begin{document}


\begin{frontmatter}

\title{
On-shell Matrix Elements of the EMT Trace in Gauge Theories and Heavy Quark Masses
}

\author[SDU]{Long Chen}
\ead{longchen@sdu.edu.cn}
\author[SDU]{Zhe Li}
\ead{lizhep@sdu.edu.cn}
\author[MPI]{Marco Niggetiedt}
\ead{marco.niggetiedt@mpp.mpg.de}

\address[SDU]{School of Physics, Shandong University, Jinan, Shandong 250100, China}
\address[MPI]{Max-Planck-Institut f\"ur Physik, Boltzmannstra{\ss}e~8, 85748 Garching, Germany}

\begin{abstract}
We present a novel diagrammatic proof of the \textit{identity} between the forward matrix element of the energy-momentum-tensor (EMT) trace operator over a single particle's on-shell state and its perturbative pole mass to any loop orders in perturbative gauge theories (with gauge bosons and fermions), without appealing to any pre-laid operator renormalization conditions or Ward identities. The proof is based on the equation of mass-dimensional analysis in dimensional regularization, the topological properties of contributing Feynman diagrams and the on-shell renormalization condition. Considering for definiteness perturbative QED and QCD with at most one fermion kept massive, we have verified the aforementioned identity, up to three loops, for all elementary particles through direct computation of dimensionally-regularized matrix elements of the relevant bare operators. Observing interestingly that the trace-anomaly contribution seemingly contains all leading-renormalon effects (explicitly verified up to three loops), we propose accordingly a new scheme- and scale-independent \textit{trace-anomaly-subtracted} mass definition for heavy $t$-,$\,b$-,$\,c$-quarks and electrons. 
A list of amusing numbers is subsequently presented for the composition of their perturbative pole masses.

\end{abstract}

\end{frontmatter}

\thispagestyle{firstpage}

\section{Introduction}
\label{sec:intro}

The classical rank-two symmetric energy-momentum tensor (EMT) describes the density and flux of energy and momentum of a physical system. 
For a closed system, its four-vector components are the conserved Noether currents associated with spacetime-translation invariance up to separately conserved terms that yield the symmetric form. 
The density and flux of energy and momentum are the sources of the gravitational field in Einstein's field equations of General Relativity. The EMT is thus indispensable in order to describe the motion of gravitating systems.
The EMT in Quantum Field Theory (QFT) has attracted substantial attention especially after the discovery of the (quantum) trace anomaly~\cite{Crewther:1972kn,Chanowitz:1972vd,Chanowitz:1972da} in the Green correlation functions between a pair of electromagnetic current operators with insertion of the trace of EMT operator. 
Appearance of this anomaly is closely connected to the broken classical scale invariance due to quantum corrections as reflected in Callan-Symanzik equations~\cite{Callan:1970yg,Symanzik:1970rt,Coleman:1970je}.
The explicit all-order (operator-level) formula for the EMT-trace operator in terms of renormalized operators has later been derived in refs.~\cite{Adler:1976zt,Collins:1976yq,Nielsen:1977sy}, where the ultraviolet (UV) finiteness of the EMT-trace operator 
in general gauge theories is fully proved along the way. 
In fact, with the aid of the general theories on renormalization of gauge-invariant composite operators~\cite{Dixon:1974ss,Joglekar:1975nu,Kluberg-Stern:1974iel,Kluberg-Stern:1974nmx,Kluberg-Stern:1975ebk,Joglekar:1976eb,Joglekar:1976pe,Henneaux:1993jn,Collins:1994ee}, ref.~\cite{Nielsen:1977sy} presented a complete and general proof that the whole EMT tensor remains UV-finite in a general quantum non-Abelian gauge theory (see also earlier refs.~\cite{Callan:1970ze,Freedman:1974gs,Freedman:1974ze}), 
one of the most celebrated properties of the \textit{total} EMT of a system in QFT.\footnote{Its individual components resulting from various feasible partitions do, in general, require respective operator renormalization, and the exact partitions of the total EMT trace thus depend also on the particular choice of individual operator renormalization, as well as regularization, schemes~\cite{Ji:1995sv,Makino:2014taa,Hatta:2018sqd,Tanaka:2018nae,Lorce:2021xku,Ahmed:2022adh}. 
Possible gauge invariant partitions of the EMT trace were discussed in ref.~\cite{Ahmed:2022adh}.}

The composition or origin of particle masses --- whether composite or not --- from the perspective of the EMT soon garnered renewed attention especially after the derivation of an explicit, all-order (operator-level) formula for the EMT trace expressed in terms of renormalized operators~\cite{Adler:1976zt,Collins:1976yq,Nielsen:1977sy}, e.~g.~in early refs.~\cite{Kashiwa:1976cp,Kashiwa:1979wc,Shifman:1978zn,Novikov:1980fa,Cheng:1991uk} and notably in refs.~\cite{Ji:1994av,Ji:1995sv} in connection with the decomposition of hadron mass (commonly referred to as the Ji's decomposition scheme).
This interest has recently been further amplified by experiments proposed at Jefferson Lab~\cite{Dudek:2012vr}, the approved Electron-Ion Collider~\cite{Accardi:2012qut}, and the currently envisaged Electron-Ion-collider in China project~\cite{Anderle:2021wcy}. 
Particular attention has been devoted to the origin and composition of the hadron mass, 
for which several alternative proposals and renewed insightful discussions have emerged~\cite{Polyakov:2002yz,Polyakov:2018zvc,Lorce:2017xzd,Hatta:2018sqd,Rodini:2020pis,Metz:2020vxd,Lorce:2021xku,Ji:2021mtz,Yang:2018nqn,He:2021bof,Kharzeev:2021qkd,Liu:2021gco,Ji:2021pys,Sun:2020ksc,Ahmed:2022adh,Czarnecki:2023yqd,Ji:2025gsq,Fujii:2025aip} (see also recent reviews~\cite{Burkert:2023wzr,Hoferichter:2025ubp} and references therein).

The fact that the expectation value of the total EMT trace for an isolated closed system should coincide with its rest mass can be expected based on general physical grounds, essentially the classical relativistic mechanics.
On the other hand, this matter is far from trivial in QFT due to quantum effects.
For example, in the seminal work~\cite{Adler:1976zt} by Adler, Collins and Duncan on the formula for the EMT trace anomaly, this equality was \textit{imposed} from the outset, i.e.~eq.~(2.6) of ref.~\cite{Adler:1976zt}, ensured via a specifically-designed yet physically natural operator subtraction condition~(2.3) therein. 
However, we have explicitly checked that these operator subtraction or  renormalization conditions are \textit{not} respected by the $\MSbar$-renormalized local operators employed in the later more systematic derivation of the trace-anomaly formula~\cite{Collins:1976yq,Nielsen:1977sy}.
A priori, it is thus not absolutely clear, at least to the authors, 
the role of the intermediate operator renormalizations in the aforementioned equality, especially in view of the subtle comments given in refs.~\cite{Adler:1976zt,Collins:1994ee} and the possible appearance of spurious technical artifacts in perturbative computations with dimensional regularization in presence of quantum anomaly\footnote{Recall, for instance, the seemingly inevitable need to call for manual corrections in the computation of axial-anomalous amplitudes at high-loop orders with $\gamma_5$ treated in dimensional regularization~\cite{Chen:2023lus,Chen:2024zju}.}. 
In principle, the aforementioned equality may be ensured by the Ward identities following from the conservation of total EMT~\cite{Callan:1970ze,Freedman:1974gs,Freedman:1974ze,Nielsen:1977sy,Kashiwa:1979wc,Fujikawa:1980rc,Caracciolo:1989pt} (with mild assumptions), typically derived using the heavy machinery of functional methods~\cite{Freedman:1974gs,Freedman:1974ze,Nielsen:1977sy,Kashiwa:1979wc,Caracciolo:1989pt}.
On the other hand, in view of the superficial differences in the contributing Feynman diagrams involved in the forward matrix elements of EMT-trace operator over single-particle states and the perturbative pole mass \textit{defined} by the zero-point of the inverse propagating function, ref.~\cite{Eides:2023uox} had recently made an effort to manifest the technical origin of such an equality for electron, especially the role of the trace-anomaly, through an explicit one-loop calculation in QED, with the aid of the explicit trace-anomaly formula~\cite{Adler:1976zt,Collins:1976yq,Nielsen:1977sy}.

In this work, we aim to consolidate the theoretical basis of the aforementioned equality between the forward matrix element of the EMT-trace operator between a single particle’s on-shell state and its perturbative pole mass 
in a transparent yet general way --- manifesting, especially, the role of trace-anomaly --- by presenting a novel direct diagrammatic proof of the equality to any loop orders in perturbative gauge theories, without appealing to any pre-laid operator renormalization conditions or Ward identities.
After proving the identity exactly for elementary fermions to arbitrary loop order in perturbative gauge theories, we perform explicit perturbative calculations, up to three loops, to show that there are indeed non-vanishing contributions from the trace-anomaly operator to the perturbative pole masses of on-shell electrons and heavy quarks; and subsequently a trace-anomaly subtracted mass 
is introduced for these elementary particles, motivated by several theoretical merits we observed in this concept. 
Besides the immediate utility in applications to perturbative high-energy physics involving heavy quarks, to be explored in future works, we hope that the insights revealed in this perturbative analysis for fundamental particles could also be of some help in gaining some better understanding or useful picture on the properties of EMT and origin of masses for non-perturbative bound states.

\section{Proof of the Identity}
\label{sec:PoI}

\subsection{The case of fermions}
\label{sec:fermioncase}

Let us begin with the fermion case, exposing a few relevant preliminaries, along which our convention and notations will be explained as well.
In perturbative QFT, the complete (connected) two-point Green's function $G_{B}$ for a fermion may be defined as (see, e.g.~\cite{Itzykson:1980rh}) 
\begin{align}\label{eq:propagatorrenormalization}
i\, G_{B}(\slashed{p}) &= \int \mathrm{d}^4\,x \, e^{+i p \cdot x} \,
 \langle 0 \big|\, \hat{\mathrm{T}} \{\psi_B(x)\, \bar{\psi}_B(0) \} \,\big| 0 \rangle_{conn.} \nonumber\\
  &= \frac{i}{\slashed{p} - m_B} \sum_{n=0}^{\infty} \Big(-i\Sigma_B(\slashed{p},\, m_B)\, \frac{i}{\slashed{p} - m_B} \Big)^n \nonumber\\
 &= \frac{i\, Z_{\psi}}{\slashed{p} - m_R - \Sigma_R(\slashed{p},\,  m_R)} \nonumber\\
&\equiv  Z_{\psi}\, i\, G_{R}(\slashed{p}) 
\end{align}
where the wave-function renormalization constant $Z_{\psi}$ encodes all net UV-divergence in $G_{B}(\slashed{p})$, such that $G_{R}(\slashed{p})$ is UV-finite. 
For the sake of simplicity in notations, the possible gauge indices of the fermion field $\psi$ (i.e.~charge and/or color indices) indicating its representation of the gauge group of the theory are all left implicit hereafter, as the form of eq.~\eqref{eq:propagatorrenormalization}, as well as the following derivation, remains essentially the same irrespective of this aspect.  
The same will be done to the gauge fields too, and the corresponding gauge (color) indices will be displayed explicitly only when they are needed for clarification.
With this definition~\eqref{eq:propagatorrenormalization}, $-i\Sigma_B$ corresponds to the expression of the Feynman amplitude for all amputated 1PI self-energy diagrams of $\psi$ field under consideration.
The dependence of $\Sigma_B$ on the bare coupling $\alpha^B$ is understood by default.
Having in mind the use of the $\MSbar$ renormalization of the bare (mass-dimensionful) coupling $\alpha^B$ in dimensional regularization (DR) with spacetime-dimension $D = 4-2\epsilon$, we follow the usual convention of defining $\alpha^B \equiv  \hat{\mu}^{2\epsilon}\, \hat{\alpha}^B  = \hat{\mu}^{2 \epsilon}\, Z_{\alpha}\, \alpha$ with the reduced bare coupling $\hat{\alpha}^B$ maintained as a mass-dimensionless quantity\footnote{Here it is unnecessary to pull out the conventional $e^{\epsilon \gamma_E} \big(4 \pi\big)^{-\epsilon}$ factor related to the usual choice of evaluating loop integrals in a particular normalization convention, which is irrelevant in our discussion.} at the expense of introducing the auxiliary mass-dimensionful variable $\hat{\mu}$ in DR (which can be set conveniently, albeit not necessarily, the same as the actual renormalization or subtraction scale $\mu$); 
on the other hand, $m_B = Z_m\, m_R$ is understood.

The bare self-energy function $\Sigma_B(\slashed{p},\, m_B)$ as defined above has a mass-dimension one.
Following from eq.~\eqref{eq:propagatorrenormalization}, one has 
\begin{equation} \label{eq:osRC_invProp}
Z_{\psi}\,\big(\slashed{p} - m_B - \Sigma_B(\slashed{p},\, m_B) \big) = 
\slashed{p} - m_R - \Sigma_R(\slashed{p},\, m_R)\,.
\end{equation} 
Within the on-shell renormalization, $G_{R}(\slashed{p})$ has a simple pole at the on-shell mass $\slashed{p} = \mOS$\footnote{With some abuse of notation, the shorthand equality $\slashed{p} = \mOS$ shall always be understood as implicitly applying to on-shell Dirac-spinors satisfying the on-shell equation of motion. 
For the sake of readers' convenience, we list a few handy equations that may be helpful to understand some of later compact derivations: $\frac{\partial }{\partial \slashed{p}} = 2\,\slashed{p} \, \frac{\partial }{\partial p^2} $ and $\slashed{p} \, \frac{\partial }{\partial \slashed{p}} =  p^{\mu}\, \frac{\partial }{\partial p^{\mu}} $.} with residue 1;
and this condition translates into the following equations in terms of the renormalized self-energy function:
\begin{equation}\label{eq:osRC_SigR}
\Sigma_R(\slashed{p} = \mOS,\, m_R=\mOS) =0\,;\quad 
\frac{\partial\, \Sigma_R(\slashed{p},\, \mOS)}{\partial\, \slashed{p}} \Big|_{\slashed{p} = \mOS} =0\,.
\end{equation} 
Therefore, $\Sigma_R(\slashed{p},\, m_R)$ in eq.~\eqref{eq:osRC_invProp} has an asymptotic series expansion around the on-shell mass $\slashed{p} = \mOS$ that begins with the quadratic power-suppression factor $\big(\slashed{p}-\mOS \big)^2$.
The same on-shell renormalization conditions can equivalently be reformulated in terms of the bare self-energy function as 
\begin{eqnarray}\label{eq:osRC_SigB}
&&\Sigma_B(\slashed{p} = \mOS,\, m_B=Z_m\, \mOS) =\big( 1 - Z_m \big)\, \mOS \,;\quad  \nonumber\\
&&\frac{\partial\, \Sigma_B(\slashed{p} = \mOS,\, m_B=Z_m\, \mOS)}{\partial\, \slashed{p}} \Big|_{\slashed{p} = \mOS} =\frac{Z_{\psi} - 1}{Z_{\psi}}\,.    
\end{eqnarray}
It is well-known that the notion of on-shell (pole) mass for an elementary fermion in a fundamental gauge theory is well-defined to \textit{any} but \textit{finite} order in perturbation theory~\cite{Bigi:1994em,Beneke:1994sw,Breckenridge:1994gs,Smith:1996xz,Kronfeld:1998di,Gambino:1999ai}. 
Therefore, the terms ``on-shell mass'' and ``perturbative pole mass'' will be used interchangeably in our discussions below.

Our starting point of the actual proof is the mass-dimensionality analysis of the dimensional-regularized bare self-energy function $\Sigma(\slashed{p}, m, \hat{\mu}) \equiv \Sigma_B(\slashed{p},\, m_B,\, \hat{\mu})$ with $\hat{\mu}$ denoting the auxiliary scale introduced in DR, which has a mass-dimension one. 
(For the sake of simplicity in notations below, we suppress several subscripts indicating ``bare'' quantities.)
Under the defining dimensional power-scaling $p \sim \lambda, \, m \sim \lambda, \, \hat{\mu} \sim \lambda, \,$ in the mass-scaling variable $\lambda$, one has the following equation of mass-dimensional analysis:
\begin{align}
\label{eq:MDAidentity}
&\lambda \frac{\mathrm{d}\, \Sigma(\slashed{p}, m, \hat{\mu})}{\mathrm{d}\, \lambda } 
=  \Sigma(\slashed{p}, m, \hat{\mu}) \nonumber\\
=\,& \lambda \frac{\mathrm{d}\, \slashed{p}}{\mathrm{d}\, \lambda } \frac{\partial\, \Sigma(\slashed{p}, m, \hat{\mu})}{\partial\, \slashed{p} } 
\, +\, \lambda \frac{\mathrm{d}\, m}{\mathrm{d}\, \lambda } \frac{\partial\, \Sigma(\slashed{p}, m, \hat{\mu})}{\partial\, m } 
\, +\, \lambda \frac{\mathrm{d}\, \hat{\mu} }{\mathrm{d}\, \lambda} \frac{\partial\, \Sigma(\slashed{p}, m, \hat{\mu})}{\partial\, \hat{\mu} } \nonumber\\
=\,& \slashed{p} \frac{\partial\, \Sigma(\slashed{p}, m, \hat{\mu})}{\partial\, \slashed{p} } 
\, +\, m \frac{\partial\, \Sigma(\slashed{p}, m, \hat{\mu})}{\partial\, m } 
\, +\, \hat{\mu} \frac{\partial\, \Sigma(\slashed{p}, m, \hat{\mu})}{\partial\, \hat{\mu} } 
\end{align} 
where the defining mass dimenisons of $\slashed{p} \sim \lambda, \, m \sim \lambda, \, \hat{\mu} \sim \lambda, \,$ and $\Sigma(\slashed{p}, m, \hat{\mu})  \sim \lambda$ are employed for deriving the first and the last equality.
We now turn to the perturbative series expansion of $m \frac{\partial\, \Sigma(\slashed{p}, m, \hat{\mu})}{\partial\, m } \, +\, \hat{\mu} \frac{\partial\, \Sigma(\slashed{p}, m, \hat{\mu})}{\partial\, \hat{\mu} }$, which, according to \eqref{eq:MDAidentity}, takes the form 
\begin{align} \label{eq:traceB}
m \frac{\partial\, \Sigma(\slashed{p}, m, \hat{\mu})}{\partial\, m } \, +\, \hat{\mu} \frac{\partial\, \Sigma(\slashed{p}, m, \hat{\mu})}{\partial\, \hat{\mu} } 
&=& \Sigma(\slashed{p}, m, \hat{\mu}) - \slashed{p} \frac{\partial\, \Sigma(\slashed{p}, m, \hat{\mu})}{\partial\, p } \,.
\end{align}
With the formal perturbative series expansion 
\begin{equation} \label{eq:SigsEXP}
\Sigma(\slashed{p}, m, \hat{\mu}) = \sum_{L=1}^{\infty} \hat{\alpha}^L\, \hat{\mu}^{(4-D)\,L}\, \Sigma^{(L)}(\slashed{p},m) = 
\sum_{L=1}^{\infty} \hat{\mu}^{2\epsilon\,L}\, \hat{\alpha}^L\,  \Sigma^{(L)}(\slashed{p},m)
\end{equation}
where the bare expansion parameter $\hat{\alpha}$ is a dimensionless quantity as declared previously, we then have 
\begin{eqnarray} \label{eq:SigsMuD}
\hat{\mu} \frac{\partial\, \Sigma(\slashed{p}, m, \hat{\mu})}{\partial\, \hat{\mu} }
&=& \sum_{L=1}^{\infty}  \hat{\alpha}^L\,  \Sigma^{(L)}(\slashed{p},m)\, 
\hat{\mu} \frac{\partial\, \hat{\mu}^{2\epsilon\,L}}{\partial\, \hat{\mu} } \nonumber\\
&=& \sum_{L=1}^{\infty} 2\epsilon\,L\, \hat{\alpha}^L\, \hat{\mu}^{2\epsilon\,L} \Sigma^{(L)}(\slashed{p},m)\,.     
\end{eqnarray} 
Now comes our key observation underlying the diagrammatic proof, which for the current case can be simply stated as follows: 
\textit{$2\epsilon\,L\, \hat{\alpha}^L\, \hat{\mu}^{2\epsilon\,L} \Sigma^{(L)}(\slashed{p},m)$ is precisely the $L$-loop bare 1PI matrix element of the operator $2\epsilon \big[-\frac{1}{4} F_{\mu\nu}\,F^{\mu\nu}\big]_B$ at zero momentum insertion between the quark states} (in Landau gauge). 
Utilizing this relation, for which a diagrammatic proof will be provided in the next subsection~\ref{sec:theorem}, and formally summing up all loops, we arrive at the following relation: 
\begin{equation}\label{eq:FFinsertion2FP} 
\langle \mathbf{p}, s \big|\, 2\epsilon \big[-\frac{1}{4} F_{\mu\nu}\,F^{\mu\nu}\big]_B \, \big| \mathbf{p}, s \rangle \big|_{\mathrm{1PI}} 
= \bar{u}(\mathbf{p}, s) \Big( \hat{\mu} \frac{\partial\, \Sigma(\slashed{p}, m, \hat{\mu})}{\partial\, \hat{\mu} }\Big) \,u(\mathbf{p}, s)     
\end{equation}
for the bare matrix element between the on-shell Dirac spinor $u(\mathbf{p}, s)$ with three-momentum $\mathbf{p}$ and helicity $s$, in accordance with our definition of the bare self-energy function in eq.~\eqref{eq:propagatorrenormalization}.

In addition, it is straightforward to see diagrammatically\footnote{by virtue of the algebraic identity $m \frac{\partial\, }{\partial\, m } \frac{i}{\slashed{p} - m} = \frac{i}{\slashed{p} - m} \big(-i \, m \big) \frac{i}{\slashed{p} - m} $}, and well-known in literature (e.g.~ref.~\cite{Lowenstein:1971jk,Adler:1976zt}), that $m \frac{\partial\, \Sigma(\slashed{p}, m, \hat{\mu})}{\partial\, m } $ gives precisely the pure loop corrections to the bare matrix element of the local fermion-mass operator $m \bar{\psi} \psi $ at zero momentum insertion between the quark states.
More specifically, we have 
\begin{equation}\label{eq:FMinsertion2FP} 
\langle \mathbf{p}, s |\, \big[m\, \bar{\psi} \psi \big]_B \, | \mathbf{p}, s \rangle \big|_{\mathrm{1PI}} 
= \bar{u}(\mathbf{p}, s) \Big( m_B \frac{\partial\,}{\partial\, m_B }\, \big(  m_B + \Sigma_B \big) \Big) \,u(\mathbf{p}, s)     
\end{equation}
in accordance with our definition of the bare self-energy function in eq.~\eqref{eq:propagatorrenormalization}.
Recall now the trace of the usual ``physical'' form of EMT, i.e.~dropping all unphysical terms not contributing to on-shell matrix elements, 
which reads in terms of the original bare operators, 
\begin{equation} \label{eq:EMTtrace}
\Theta^{\mu}_{\mu} \big|_{\mathrm{phy.}} = 2\epsilon \big[-\frac{1}{4}  F^{a}_{\rho\sigma}\,F^{a\, \rho\sigma}\big]_B + \big[m \bar{\psi} \psi\big]_B \,,
\end{equation}
in the case of only one flavor of fermion kept massive in the gauge theory.
With a slight abuse of notations, we use $F^{a}_{\mu \nu}\,F^{a\, \mu \nu}$ with $a =1, \cdots, N_c^2-1$ for the squared gauge-field strength tensor in a non-Abelian SU($N_c$)-gauge theory (with QCD as a prototype), and formally the same notation with $a=0$ for the squared gauge-field strength tensor in an Abelian gauge theory (with QED as a prototype).
In view of the aforementioned points, it should be evident now the next step of the demonstration is to rewrite the combination $Z_{\psi}\, \big(m + \Sigma(\slashed{p}, m, \hat{\mu}) - \slashed{p} \frac{\partial\, \Sigma(\slashed{p}, m, \hat{\mu})}{\partial\, \slashed{p} } \big)$ at on-shell kinematics in terms of the on-shell renormalized quantities by virtue of on-shell renormalization condition~\eqref{eq:osRC_SigR}: 
\begin{eqnarray} \label{eq:osMComposition}
&& Z_{\psi}\, \Big(m +  m \frac{\partial\, \Sigma(\slashed{p}, m, \hat{\mu})}{\partial\, m } \, +\, \hat{\mu} \frac{\partial\, \Sigma(\slashed{p}, m, \hat{\mu})}{\partial\, \hat{\mu} }  \Big) \Big|_{\slashed{p}=\mOS} \nonumber\\
&=&
Z_{\psi}\, \Big(m + \Sigma(\slashed{p}, m, \hat{\mu}) - \slashed{p} \frac{\partial\, \Sigma(\slashed{p}, m, \hat{\mu})}{\partial\, \slashed{p} } \Big) \Big|_{\slashed{p}=\mOS}  \nonumber\\
&=& 
Z_{\psi}\, \Big(m + \Sigma(\slashed{p}, m, \hat{\mu}) - \slashed{p} \frac{Z_{\psi}-1}{Z_{\psi}} \Big) \Big|_{\slashed{p}=\mOS} \nonumber\\
&=& 
\Big( \big( \mOS + \Sigma_R(\slashed{p}, \mOS, \hat{\mu}) - \slashed{p} \big) + \slashed{p} \Big) \Big|_{\slashed{p}=\mOS} \nonumber\\
&=&  \mOS 
\end{eqnarray} 
where we have employed the on-shell renormalization condition~\eqref{eq:osRC_SigB} $\frac{\partial\, \Sigma(\slashed{p}, m, \hat{\mu})}{\partial\, \slashed{p} } \big|_{\slashed{p}=\mOS} = \frac{Z_{\psi}-1}{Z_{\psi}}$ for the bare self-energy function evaluated at the on-shell point corresponding to the pole mass $\slashed{p}=\mOS$ and $\Sigma_R(\slashed{p}, \mOS, \hat{\mu})\big|_{\slashed{p}=\mOS} = 0$ for the renormalized self-energy function.
Thus we have demonstrated that the equality holds exactly with bare operators (without any operator renormalization) to any loop orders and also to all power orders in the DR regulator $\epsilon$ (i.e.~exactly with the DR-regulator kept without taking the 4-dimensional limit), but only if the on-shell renormalization condition is used for the external states.

When there are several massive fermions engaged in the gauge interactions, there will be several logarithmic derivatives with respect to the fermion masses in the first line of \eqref{eq:osMComposition}, whereas the form of the second line remains essentially the same.  
It is straightforward to see that these additional pieces are in one-to-one correspondence with the additional fermion-mass operators contained in the total EMT trace in the presence of multiple massive fermions (see, e.g.~eq.~\eqref{eq:EMTtrace_GF}).
Consequently the above demonstration remains valid in gauge theories irrespective of the presence of additional fermions.

In terms of the matrix element evaluated with the on-shell spinor $u(\mathbf{p}, s )$, our result~\eqref{eq:osMComposition} can be formulated as 
\begin{equation}\label{eq:OSmassIdentity}
\langle \mathbf{p}, s  \big|\, 2\epsilon \big[-\frac{1}{4}  F^{a}_{\rho\sigma}\,F^{a\, \rho\sigma}\big]_B + \big[m \bar{\psi} \psi\big]_B  \, \big| \mathbf{p}, s \rangle = \bar{u}(\mathbf{p}, s ) \, \mOS \, u(\mathbf{p}, s)     
\end{equation}
which holds as an \textit{identity} for any $u(\mathbf{p}, s)$ independent of its helicity and normalization convention. 
In our practical computations, we employed the following trace technique: 
$$\sum_{s = \pm} \bar{u}(\mathbf{p}, s ) \, \Gamma \,  u(\mathbf{p}, s ) = \mathrm{Tr}\Big[\Gamma\, \big(\slashed{p} + m \big) \Big] $$
and adopted the usual relativistic normalization convention for a single particle state.
\footnote{Note that $\bar{u}(\mathbf{p}, s ) \, \Gamma \,  u(\mathbf{p}, s ) = \mathrm{Tr}\Big[\Gamma\, \big(\slashed{p} + m \big) \, \big(1 + \gamma_5 \slashed{s} \big)/2 \Big] $ which can be further reduced to $\mathrm{Tr}\Big[\Gamma\, \big(\slashed{p} + m \big)/2\Big]$ if $\Gamma$ is free of $\gamma_5$, such as in the chiral-symmetric gauge theories in consideration.}
Our demonstration shows that with this relativistic state normalization, the above relation~\eqref{eq:OSmassIdentity} between $\langle \mathbf{p}, s \big|\, \Theta^{\mu}_{\mu} \big|_{\mathrm{phy.}}  \, \big| \mathbf{p}, s \rangle$ and $\bar{u}(\mathbf{p}, s) \, \mOS \, u(\mathbf{p}, s) = 2 \, m^2_{os}$ is a mathematical \textit{identity}, rather than an \textit{equation}, in perturbative quantum gauge theories. 

\begin{figure}[htbp]
\centering
\includegraphics[width = 0.46\textwidth]{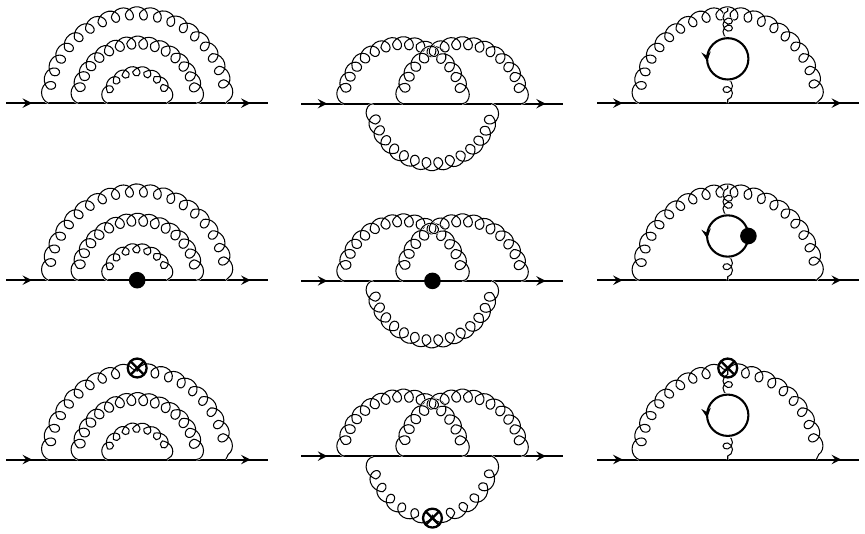} 
\caption{The first row of sample diagrams is for the self-energy function of a massive fermion at three loops; the second row represents the counterpart in the contribution to the matrix element of the fermion-mass operator $m \bar{\psi} \psi$ (with the black dot denoting the vertex factor $i\, m$); 
the last row of example diagrams illustrates some of the three-loop contribution with an insertion of $\mathcal{O}_{F}[\xi]$, indicated by the circled cross, as defined in eq.~\eqref{eq:EMTtrace_GF}. 
}
\label{fig:RepFeynDiags}
\end{figure}

To be on the safe side, especially in view of some subtleties for local composite operators at the zero-momentum insertion~\cite{Joglekar:1976eb,Collins:1994ee}, we have explicitly computed the matrix elements of the two independent local composite operators involved in $\Theta^{\mu}_{\mu} \big|_{\mathrm{phy.}} = 2\epsilon \big[-\frac{1}{4}  F^{a}_{\rho\sigma}\,F^{a\, \rho\sigma}\big]_B + \big[m \bar{\psi} \psi\big]_B \,,$ between a pair of on-shell fermions (and between a pair of gauge bosons, as will be discussed later), up to three loops, and subsequently conducted an explicit verification of the above identity up to this perturbative order. 
For illustration, several representative Feynman diagrams involved in the three-loop calculations are shown in figure~\ref{fig:RepFeynDiags}. 
On the other hand, there are a few interesting practical outcomes resulting from such explicit three-loop calculations, which will be reported in  section~\ref{sec:OSMcomp_TASmass} and~\ref{ApxSec:Z3Z3res}.

\subsection{A theorem on insertion of $\mathcal{O}_{F}[\xi]$ into vacuum diagrams}
\label{sec:theorem}

Consider a local gauge quantum field theory with only gauge bosons and fermions, we will prove the following theorem:
\textit{the sum of all L-loop 1PI vacuum diagrams with an insertion of $\mathcal{O}_{F}[\xi] \equiv \big[-\frac{1}{4}  F^{a}_{\mu \nu}\,F^{a\, \mu \nu} -\frac{1}{2 \xi} \big(\partial_{\mu} A^{\mu}_a \big)^2 \big]_B$ at zero momentum transfer, i.e.~the 1PI vacuum-vacuum matrix element of $\mathcal{O}_{F}[\xi]$, is proportional to the sum of all original L-loop 1PI vacuum diagrams, at the loop-integrand level, with an overall factor $L-1$ for $L \geq 2$}. 
As mentioned before, slightly abusing the notations, we use $F^{a}_{\mu \nu}\,F^{a\, \mu \nu}$ with $a =1, \cdots, N_c^2-1$ to denote the squared gauge-field strength tensor in a non-Abelian SU($N_c$)-gauge theory (with QCD as a prototype), and formally the same notation with $a=0$ for an Abelian gauge theory (with QED as a prototype).

Upon eliminating the unphysical terms vanishing by applying equation-of-motion on external (on-shell) states and the total derivative term, 
the trace of the full EMT tensor with the gauge-fixing term reads~\cite{Nielsen:1977sy}
\begin{equation} \label{eq:EMTtrace_GF}
\Theta^{\mu}_{\mu} 
= 2\epsilon\, \mathcal{O}_{F}[\xi]  + \sum_f \big[m_f \bar{\psi}_f \psi_f\big]_B \,,    
\end{equation}
where the gauge-dependent term $2\epsilon \big[-\frac{1}{2 \xi} \big(\partial_{\mu} A^{\mu}_a \big)^2 \big]_B$ contained in $2\epsilon\, \mathcal{O}_{F}[\xi]$ is kept for convenience --- to have the asserted diagrammatic relation manifestly valid in any $\xi$ gauge without taking the on-shell kinematics --- which does not eventually contribute to the ($\xi$-independent) on-shell renormalized matrix elements.\footnote{Note, however, both the bare on-shell S-matrix elements of the operators in EMT-trace and the $Z_{\psi}^{os}$ depend on $\xi$, although the latter starts to appear only from three loops~\cite{Melnikov:2000zc}; more comments on this will be given later in~\ref{ApxSec:Z3Z3res}.} 
Note that the unphysical gauge-fixing term in eq.~\eqref{eq:EMTtrace_GF}, i.e.~$2\epsilon \big[-\frac{1}{2 \xi} \big(\partial_{\mu} A^{\mu}_a \big)^2 \big]_B$ effectively vanishes in the Landau-gauge $\xi = 0$.

\begin{figure}[htbp] 
\centering
\includegraphics[width = 0.48\textwidth]{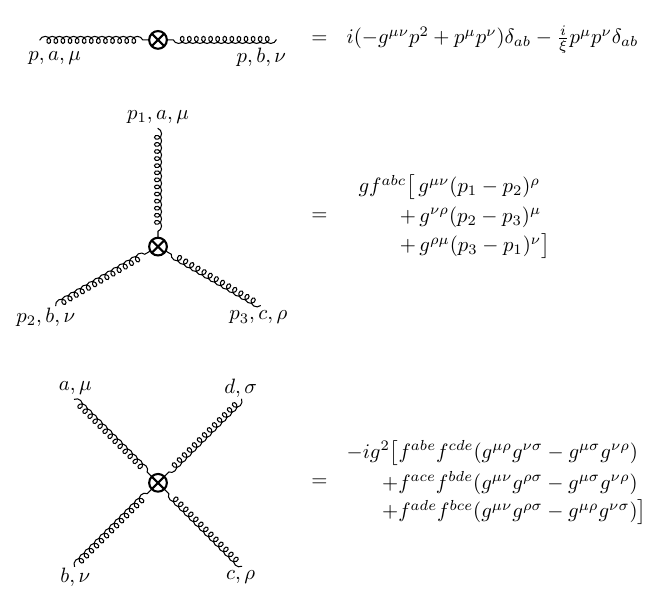} 
\caption{The Feynman rules for the local composite operator $\mathcal{O}_{F}[\xi]$ with zero-momentum insertion, represented by the circled cross (all momenta are outgoing). 
}
\label{fig:FFoperator_FeynRule}
\end{figure}

\begin{figure}[htbp]
\centering
\includegraphics[width = 0.45\textwidth]{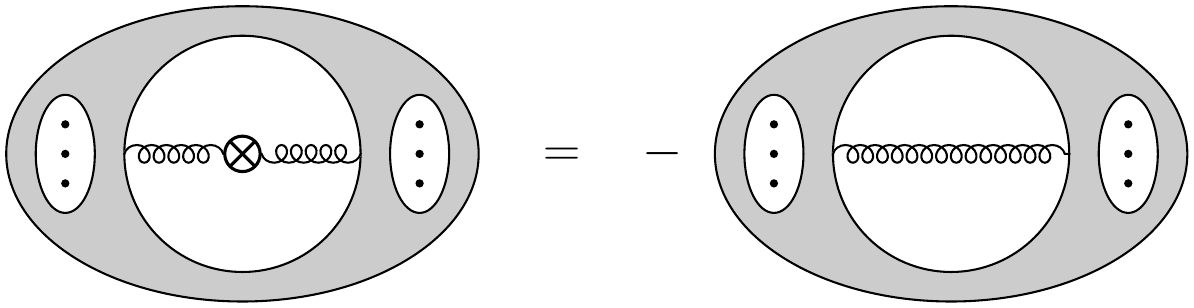} 
\caption{The reduction of the insertion of a degree-2 vertex onto an internal gauge boson propagator inside a generic diagram (note the relative minus sign in the r.h.s.).}
\label{fig:propagatorinsertion}
\end{figure}

The Feynman rules for the local composite operator $\mathcal{O}_{F}[\xi]$ with zero-momentum insertion can be readily derived following the standard textbook procedure.
They are documented in figure~\ref{fig:FFoperator_FeynRule} for the convenience of later reference.
In particular, the Feynman rule for the degree-2 vertex of $\mathcal{O}_{F}[\xi]$ with zero momentum insertion 
reads 
$$
i\, \big(-g^{\mu\nu}\,p^2 + (1-1/\xi) \, p^{\mu} p^{\nu}  \big)\, \delta_{a b}\,.
$$
It is worth noting, which is crucial for the following discussions, that this expression happens to be the \textit{minus} inverse of the gauge propagator in a general $\xi$-gauge,  
$$i\frac{-g^{\mu\nu} + (1 -\xi) \frac{p^{\mu} p^{\nu}}{p^2}}{p^2 + i \epsilon_0^{+}}  \delta_{ab} \,.$$
Consequently, the insertion of such a degree-2 vertex onto any gauge boson propagator yields a single boson propagator but with an \textit{opposite} sign, as illustrated diagrammatically in figure~\ref{fig:propagatorinsertion}.

On the other hand, the Feynman rules for the degree-3 and degree-4 vertices of $\mathcal{O}_{F}[\xi]$, which are listed in figure~\ref{fig:FFoperator_FeynRule}, are  identical to the triple and quartic gauge-boson self-coupling vertices in gauge theories. 
Apparently, the quadratic gauge-fixing term $\,-\, \frac{1}{2 \xi} \big(\partial_{\mu} A^{\mu}_a \big)^2$ gives only a degree-2 vertex; 
the degree-3 and degree-4 vertices appear only in non-Abelian gauge theories and are the same as those for the local-composite operator $-\frac{1}{4} F^{a}_{\rho\sigma}\,F^{a\, \rho\sigma} $.

Given the above insights, in particular the illustration in figure~\ref{fig:propagatorinsertion}, it shall now be evident that the net result of inserting the aforementioned vertices in figure~\ref{fig:FFoperator_FeynRule} into a given vacuum Feynman diagram is determined by the number of gauge-boson propagators, minus the number of the triple and quartic gauge-boson self-coupling vertices, at which the operator $\mathcal{O}_{F}[\xi]$ can be attached. 
Let us denote by $N^{\mathrm{eff}}_{g}$ the number of the \textit{effective gauge-boson propagators} with non-vanishing attachment of $\mathcal{O}_{F}[\xi]$.
For a given 1PI vacuum diagram, let  $N_g$ and $N_f$ denote, respectively, the number of the (internal) gauge boson and fermion propagators, and $V_f$, $V_3$ and $V_4$ for the number of fermion-gauge, triple-gauge and quartic-gauge coupling vertices therein, respectively.
Then we have 
\begin{equation} \label{eq:Ngeff_def}
N^{\mathrm{eff}}_{g} = N_g - V_3 - V_4\,.     
\end{equation}
Following from the topological property of a linear graph, essentially every internal line joining exactly two internal vertices, we have for this vacuum diagram 
\begin{equation} \label{eq:Ngeff}
2 \, \big( N_g + N_f \big) = 3\, V_f \,+\, 3\, V_3 \,+\, 4\, V_4 \,,    
\end{equation}
as there are no external legs, and the number of loops given by 
\begin{equation} \label{eq:NLoops}     
N_L = \big( N_g + N_f \big) - \big( V_f  + V_3 + V_4 \big)  \,+\, 1 \,.
\end{equation}
Combining the above relations, we then obtain the following formula for $N^{\mathrm{eff}}_{g}$, the number of effective gauge-boson propagators in a $N_L$-loop vacuum diagram: 
\begin{eqnarray} \label{eq:Ngeff_to_NL}
N^{\mathrm{eff}}_{g} &=&  \big( V_f  + V_3 + 2\, V_4 \big) /2 \nonumber\\
&=& N_L - 1\, 
\end{eqnarray}
where we have employed, in addition, a topological relation $N_f = V_f$ for every fermion propagator is part of a fermion loop in the vacuum diagram 
(i.e.~there is no open fermion chain).\footnote{The incorporation of the Faddeev-Popov ghost fields in the Feynman rules of the gauge theory, absent in certain gauge-fixing conditions, does not change this final result~\eqref{eq:Ngeff_to_NL}. 
}
Consequently, this completes the theorem stated at the beginning of this subsection on the insertion of $\mathcal{O}_{F}[\xi]$ into vacuum diagrams in perturbative gauge theories, including both Abelian and non-Abelian cases.
~\\

To go from the number $N^{\mathrm{eff}}_{g}$ of effective gauge-boson propagators for a vacuum diagram to a forward-scattering diagram with external legs, such as the fermion or gauge boson propagators, simply cut open a \textit{fixed} number of propagators corresponding exactly to the external legs of the diagram in question and note further that \textit{no} more attachment of $\mathcal{O}_{F}[\xi]$ to the gauge propagators being cut open.\footnote{In principle, the number of effective gauge-boson propagators defined in eq.~\eqref{eq:Ngeff_def} for a generic Feynman diagram may also be derived directly without going via vacuum diagrams. Consider a connected Feynman diagram with $E_F$ fermion legs and $E_B$ gauge boson leg, and then by essentially eq.~\eqref{eq:NLoops} and the generalized $N_f = V_f - E_F/2$, the result reads $N^{\mathrm{eff}}_{g} = N_L - 1 + E_F/2$ independent of $E_B$.}

Now let us consider, in particular, the case of fermion-propagator diagrams addressed in the preceding subsection, and supply the link between eq.~\eqref{eq:SigsMuD} and eq.~\eqref{eq:FFinsertion2FP}.
Any Feynman diagram representing $L$-loop corrections to the self-energy of a fermion can be obtained or viewed as a corresponding $L+1$-loop vacuum diagram with exactly one fermion loop being cut open.
Consequently, the number of the effective gauge-boson propagators in any $L$-loop fermion self-energy correction diagram equals $L$.
Based on this information, and noting that the amputated quark-quark matrix element $\langle \mathbf{p}, s \big|\, \big[-\frac{1}{4}  F^{a}_{\rho\sigma}\,F^{a\, \rho\sigma} \big]_B  \, \big| \mathbf{p}, s  \rangle \big|_{\mathrm{1PI}}$ is evaluated technically as $-i$ times the 1PI Feynman amplitude (to comply with the same convention adopted in the evaluation of the fermion-mass operator $m \bar{\psi} \psi$),  
it is then straightforward to see that for the $L$-loop contribution, 
$$\langle \mathbf{p}, s \big|\, \big[ -\frac{1}{4}  F^{a}_{\rho\sigma}\,F^{a\, \rho\sigma} \big]_B \,  \big|\mathbf{p}, s \rangle \big|_{\mathrm{1PI}}^{L} = L\, \bar{u}(\mathbf{p}, s)\, \Sigma_{B} \,u(\mathbf{p}, s) \big|_{\mathrm{1PI}}^{L}\,$$
holds. 
We have thus completed the demonstration of the missing link between eq.~\eqref{eq:SigsMuD} and eq.~\eqref{eq:FFinsertion2FP}, and hence the proof intended in the previous subsection.

\subsection{The case of gauge bosons}
\label{sec:gaugecase}

Gauge bosons remain exactly massless at any perturbative order in perturbative gauge theories with exact or unbroken gauge symmetry.
With the aid of the above theorem and the transversality of the standard gauge-boson self-energy or vacuum-polarization tensor 
$$\Pi^{\mu\nu} (p, m, \hat{\mu}) = \big(-g^{\mu\nu}\,p^2 + p^{\mu} p^{\nu} \big)\, \Pi_0(p, m, \hat{\mu})\,,$$ 
it is even simpler to show the vanishing of the matrix element of $\Theta^{\mu}_{\mu}$ between two on-shell gauge boson states, given that the dimensionless form factor $\Pi_0$, known as the transversal vacuum-polarization function, has a limit at $p^2=0$.
Indeed, this limit determines precisely the on-shell gluon wave-function renormalization constant, whose explicit perturbative expression up to three loops with gauge parameter dependence will be presented in~\ref{ApxSec:Z3Z3res}.

Any Feynman diagram representing $L$-loop corrections to the self-energy of a gauge boson can be obtained or viewed as a corresponding $L+1$-loop vacuum diagram with exactly one gauge boson propagator being cut open.
Consequently, the number of the effective gauge-boson propagators in any $L$-loop gauge boson self-energy correction diagram remains $L-1$ as in eq.~\eqref{eq:Ngeff_to_NL}, as both the number of loops and that of allowed operator attachments are reduced, respectively, by one due to this cut.
Therefore, the $L$-loop gluon-gluon matrix element $\langle \mathbf{p}, s \big|\, \big[ -\frac{1}{4}  F^{a}_{\rho\sigma}\,F^{a\, \rho\sigma} \big]_B  \, \big|\mathbf{p}, s \rangle \big|_{\mathrm{1PI}}^{L} $ is proportional to $ \big( L-1 \big)\, \Pi^{\mu\nu} \,\big|_{\mathrm{1PI}}^{L}$, and thus vanishes at $p^2=0$ provided the aforementioned existence of the on-shell gluon wavefunction renormalization constant.

As for the contribution from the fermion-mass operator $m \bar{\psi} \psi$ in the EMT trace $\Theta^{\mu}_{\mu}$, it is obvious that a formula similar to  eq.~\eqref{eq:FMinsertion2FP} holds here as well.
Therefore, it is proportional to the logarithmic derivative of the vacuum-polarization tensor $\Pi^{\mu\nu}$ with respect to the fermion mass.
The vanishing of such a term at $p^2=0$ is ensured by the existence of the logarithmic derivative of the on-shell gluon wave-function renormalization constant with respect to fermion mass. 
We have thus demonstrated that the matrix element of $\Theta^{\mu}_{\mu}$ between two on-shell gauge boson states vanishes at on-shell kinematics, in compliance with gauge bosons being exactly massless in perturbative gauge theories.
We emphasize that, in the cases of on-shell gauge bosons, the contributions from the trace-anomaly operator and the fermion-mass operator vanish respectively, i.e.~independent of each other.
~\\

Alternatively, if one combines the two pieces in the matrix element of $\Theta^{\mu}_{\mu}$ at \textit{off-shell momentum} $p$, by repeating some of the derivations in subsection~\ref{sec:fermioncase} together with the knowledge of the proportionality factor $N^{\mathrm{eff}}_{g} = L-1$ for gauge-boson self-energy diagrams, 
it then becomes rather straightforward to arrive at 
$$
\big(2 - p^{\,\rho} \frac{\partial}{ \partial p^{\,\rho}} \big)\, \Pi^{\mu\nu} = 
\big(g^{\mu\nu}\, p^2  - p^{\mu} p^{\nu} \big) \, 2 p^2\, \frac{\mathrm{d}\, \Pi_0}{\mathrm{d}\, p^2}
$$
for the off-shell matrix element.
This result is known as the ``(quantum) trace identities'' studied, e.g.~in the early refs.~\cite{Callan:1970ze,Chanowitz:1972vd,Chanowitz:1972da} and systematically re-analyzed later in refs.~\cite{Freedman:1974gs,Freedman:1974ze,Adler:1976zt,Collins:1976yq,Nielsen:1977sy,Kashiwa:1979wc} (see also similar discussions in refs.~\cite{Caracciolo:1989pt,Caracciolo:1991cp,Caracciolo:1991vc}).
It is clear that the vanishing of the matrix element of $\Theta^{\mu}_{\mu}$ between two on-shell gauge boson states at the on-shell point may be expected using this result as well, provided the existence of $p^2\, \frac{\mathrm{d}\, \Pi_0}{\mathrm{d}\, p^2}$, sometimes called Adler function, at the on-shell point. 
~\\

It is  straightforward to see that the analysis carried out in the previous subsections~\ref{sec:fermioncase} and~\ref{sec:theorem}, in particular eq.~\eqref{eq:MDAidentity} and eq.~\eqref{eq:osMComposition} as well as the derivation of the effective number $N^{\mathrm{eff}}_{g}$, can be generalized to more general cases with $E_F$ fermion legs and $E_B$ gauge boson legs.
This can readily allow us to make direct assertions regarding the matrix elements of the local operator $\mathcal{O}_{F}[\xi] \equiv \big[-\frac{1}{4}  F^{a}_{\mu \nu}\,F^{a\, \mu \nu} -\frac{1}{2 \xi} \big(\partial_{\mu} A^{\mu}_a \big)^2 \big]_B$ as well as $\Theta^{\mu}_{\mu} $ between various external quantum states considered in some phenomenological scattering process.
We leave this to future works.

\section{Trace-anomaly subtracted mass for fermions in perturbative QED and QCD}
\label{sec:OSMcomp_TASmass}

The exact identity~\eqref{eq:OSmassIdentity} for elementary fermions proved in the subsections~\ref{sec:fermioncase} and ~\ref{sec:theorem} justifies the subsequent discussion of the composition of the on-shell masses for electron and heavy quarks, which are well-defined to any finite order~\cite{Bigi:1994em,Beneke:1994sw,Breckenridge:1994gs,Smith:1996xz,Kronfeld:1998di,Gambino:1999ai} in perturbative QED and QCD. 
The two respectively UV-finite and gauge-invariant bare local operators therein naturally suggest a partition of the perturbative quantum corrections encoded in the self-energy function which brings the mass of a fermion from the bare value to the on-shell perturbative pole mass in gauge theories, which is obvious from eq.~\eqref{eq:osMComposition}: 
i) the ``classical'' fermion-mass operator part, whose origin within the scope of the Standard Model is the Yukawa interaction with the Higgs field with non-vanishing vacuum condensation, which may be simply called the Higgs-generated mass contribution; 
ii) the other part is fully given by the matrix element of the bare trace-anomaly operator $2\epsilon\, \mathcal{O}_{F}[\xi]$ which reflects the inevitable presence of the UV-\textit{regulator} effect signifying the technical breaking of classical scale symmetry in QFT. 
Apparently, the analysis of the perturbative pole masses of elementary fermions from the perspective of EMT trace proceeds in parallel to the much more involved discussions on the structure of hadron masses alluded to in the Introduction, except for the fact that the matrix elements involved in our case can all be computed perturbatively to any finite order.

Working in the perturbative QCD with one massive quark and $n_l$ massless quark flavors, we have obtained the following analytic three-loop result for the portion of the Higgs-generated mass contribution to the perturbative pole mass of the heavy quark:
{
\begin{align} \label{eq:TASmass_OSmass}
\mathrm{Z}_{\sigma} \equiv&  \frac{\langle \mathbf{p}, s  \big| \big[m \bar{\psi} \psi\big]_B  \, \big| \mathbf{p}, s \rangle}{\bar{u}(\mathbf{p}, s ) \, \mOS \, u(\mathbf{p}, s) } \equiv 1 - \ZTA \nn
={}&
1 - \api \frac{3 C_F}{2} 
+ \Bigg(\api \Bigg)^2 \frac{C_F}{48}  
\Big[ 
-185 C_A + 99 C_F 
\nn
& 
+ 26 (n_h + n_l) 
+ \big( -66 C_A + 12 (n_h + n_l) \big) \loguos 
\Big] 
\nn 
& 
- \Bigg(\api \Bigg)^3 \frac{C_F}{1728}  
\Bigg\{ 
8343 C_F^2 + 621 C_F( n_h + n_l) 
+ 788 n_h^2 
\nn
&
+ 356 n_l^2
+ 1144 n_h n_l 
 - 360 C_F\pi^2( n_h + n_l)
- 96 n_h^2 \pi^2 
\nn
&
 + 48 n_l^2 \pi^2 
- 48 n_h n_l \pi^2
+ 576 C_F\pi^2 \log(2)  (n_h +n_l) 
\nn
&
+ 6 \Big[ 2341 C_A^2 
+ 4 (n_h + n_l) \big( 36 C_F + 13 (n_h + n_l) \big) 
\nn
&
- C_A \big( 1089 C_F
 + 746 (n_h + n_l) \big) 
 \Big] \loguos 
+ 18 \big( 11 C_A 
\nn
&
- 2 (n_h + n_l) \big)^2 \loguos^2 
+ C_A^2 \Big[ 26486 
+ 528 \pi^2 (-1 + 3\log(2)) 
\nn
&
- 2376 \zeta_3 
\Big]
+ 432 C_F(n_h + n_l)  \zeta_3  
- 4 C_A \Big[ 
9 C_F \big( 525
\nn
&
 + 11 \pi^2 (-5 
+ 8\log(2)) 
- 132 \zeta_3 \big) 
+ 4 n_h \big( 583 + 3 \pi^2 (-13 
\nn
&
+ 6\log(2) ) 
+ 54 \zeta_3 \big) 
+ 2 n_l \big( 869 + 3 \pi^2 (7 + 12\log(2))) 
\nn
&
+ 108 \zeta_3 \big) 
\Big] 
\Bigg\}
\end{align}
}
where $\loguos \equiv \ln\big(\mu^2/\mOS^2 \big)$ denotes the logarithm of the squared ratio of the scale $\mu$ of the $\MSbar$-renormalized QCD coupling $\alpha_s$ over the perturbative pole mass $\mOS$ of the heavy quark.
$C_F\,,C_A$  are the standard Casimir gauge-group invariants or color factors, which in the case of QCD read $C_A=3\,, C_F=4/3$.
We have also introduced a handy tag $n_h$ per massive fermion loop, which shall be set to 1 to comply with the fact that the theory has only one fermion kept massive, for the sake of extracting the corresponding result in Abelian gauge theory\footnote{We note that the formula given in ref.~\cite{Adler:1976zt} for electron cannot be directly applied here, due to different intermediate renormalization conditions employed.}.
More specifically, the result in QED up to three loops can be derived from eq.~\eqref{eq:TASmass_OSmass} with the usual replacement rule:  $C_A \rightarrow 0,\, C_F \rightarrow 1,\, n_l \rightarrow 2\, n_l,\, n_h \rightarrow 2\, n_h$.

According to eq.~\eqref{eq:OSmassIdentity}, $\SigmaMass \equiv \mathrm{Z}_{\sigma} \, \mOS $ may be viewed as a residual Higgs-generated mass for an on-shell fermion obtained by subtracting away the trace-anomaly contribution from its perturbative pole mass.\footnote{In view of the terminology employed in the parallel discussion on the origin and structure of hadron masses and in Lattice QCD calculations, one may call it the $\sigma$-mass of heavy quarks.}
Obviously, $\mathrm{Z}_{\scriptscriptstyle{\mathrm{TA}}} = 1-\mathrm{Z}_{\sigma}$ corresponds to the non-vanishing portion taken up by the trace-anomaly contribution to the perturbative pole masses of elementary fermions, which we also computed directly in order to explicitly verify eq.~\eqref{eq:OSmassIdentity} up to three loops.  
Why would this particular mass definition be of any interest for application to scattering processes involving heavy quarks in perturbative QCD?
It is straightforward to see that $\SigmaMass$ is gauge-invariant, regularization/renormalization scheme- and scale-independent, because the bare Higgs-generated fermion-mass operator $\big[m \bar{\psi} \psi\big]_B$ is itself a UV-finite gauge-invariant object and the zero-point of the inverse propagator (i.e.~the pole position of the propagator to identify the on-shell asymptotic state) is also scheme- and gauge-invariant~\cite{Breckenridge:1994gs,Smith:1996xz,Kronfeld:1998di,Gambino:1999ai}. 
On top of this, via examining the perturbative relation of $\SigmaMass$ to the renormalon-free $\MSbar$ mass, 
which is documented in eq.\eqref{eq:TASmass_MSmass} in~\ref{ApxSec:addmassrelations}\footnote{For the sake of readers' convenience, a supplemental file containing the electronic form for all analytic expressions reported in this work is provided.}, we find, with a bit of surprise, that this perturbative relation is free of the characteristic leading IR-renormalon behavior~\cite{Bigi:1994em,Beneke:1994sw,Beneke:1998ui} in the large-$n_f$ approximation (or ``naive non-abelianization'')~\cite{Beneke:1994qe,Broadhurst:1994se}. 
In other words, we observe that the trace-anomaly contribution nicely contains \textit{all} leading IR-renormalon effects in the definition of the perturbative pole mass, at least explicitly verified up to three loops. 
(A more detailed analysis on this point will be presented in a forthcoming work~\cite{Chen:2025zfa}.)
Given the aforementioned appealing features, we thus propose to take $\SigmaMass \equiv \mathrm{Z}_{\sigma} \, \mOS $ as a new scheme- and scale-invariant (leading-anomaly-free) mass definition for the electron and heavy quarks in perturbative gauge theories, which we call \textit{trace-anomaly-subtracted mass}.
In particular, this mass definition for electron and heavy quarks happens to combine the merits of both the on-shell mass and the $\MSbar$ mass definition, while elegantly circumvents their respective unappealing and undesirable features.
We note that, in addition to the perturbative pole mass and $\MSbar$-mass, there are several useful alternative short-distance mass definitions of heavy quarks proposed in the literature, each motivated by distinct theoretical or practical considerations;
an incomplete list includes the kinetic mass~\cite{Bigi:1994ga,Bigi:1996si,Czarnecki:1997sz,Fael:2020iea}, the potential-subtracted mass~\cite{Beneke:1998rk}, the 1S-mass~\cite{Hoang:2008yj}, the MSR-mass~\cite{ Hoang:1998ng,Hoang:2017suc}, the (minimal) renormalon-subtracted mass~\cite{Pineda:2001zq,Komijani:2017vep}, the RI/MOM mass~\cite{Martinelli:1994ty} and RI/(m)SMOM mass~\cite{Aoki:2007xm,Sturm:2009kb,Boyle:2016wis}.
When needed, the perturbative relations between $\SigmaMass$ and these masses can be readily derived up to three loop orders using eq.~\eqref{eq:TASmass_OSmass} and eq.~\eqref{eq:TASmass_MSmass}, provided their relationships to the on-shell or $\MSbar$ masses are known at least to the same order (which are mostly the case now, see, e.g.~the recent comprehensive review~\cite{Beneke:2021lkq} and the compilations in refs.~\cite{Chetyrkin:2000yt,Herren:2017osy}).

To quickly get some more concrete ideas about this mass, we provide, in table~\ref{tab:tasmasses}, our numerical results for the portions occupied by the trace-anomaly contributions to the perturbative pole masses of the electron, $t$-quark, $b$-quark and $c$-quark, as well as the corresponding trace-anomaly-subtracted $\sigma$-masses, determined using the three-loop relation~\eqref{eq:TASmass_OSmass} and \eqref{eq:TASmass_MSmass} in perturbative QCD and QED.
\begin{table}[htbp]
\centering
\begin{tabular}{ c  c  c  c c } \\
\toprule[1.2pt]
 & electron & $t$-quark & $b$-quark & $c$-quark \\
\midrule
\midrule
$\ZTA$ & 0.347\,\% & 7.9\,\% & 20.4\,\% & 34.3\,\% \\
\midrule
$\SigmaMass$ & $0.509$~MeV & $159.0$~GeV & $3.96$~GeV & $1.17$~GeV \\
\bottomrule[1.2pt]
\end{tabular} 
\caption{The numbers in the second row are for the portion $\ZTA$ taken up by the trace-anomaly contributions to the perturbative pole masses of the electron, $t$-quark, $b$-quark and $c$-quark; 
the bottom row lists the values of the corresponding trace-anomaly-subtracted $\sigma$-masses.
(Further details on these numbers are provided in the text)}
\label{tab:tasmasses}
\end{table}
Rather than aiming to deliver the final numbers that would result only from a fully-fledged analysis keeping track of all sorts of systematic uncertainties and errors in (experimental) input parameters, which will definitely be improved over time, we focus here, for the moment, on providing the central values that are sufficiently accurate to illustrate the main features and interesting patterns exhibited in the $\SigmaMass$ values determined for these particles.

A few general comments on the numbers in table~\ref{tab:tasmasses} are now in order.
The portion $\ZTA$ and $\SigmaMass$ are, in principle, independent of the $\alpha_s$-renormalization scale $\mu$, a property that holds order-by-order in perturbation theory. 
In the approximation of only one active heavy quark $Q$ kept massive in the effective QCD Lagrangian,\footnote{The effect of another massive quark enters only virtually and starts from two-loop corrections. With an appropriate choice of the low-energy effective QCD Lagrangian with the heavier quark fields \textit{integrated out} and lighter quark fields approximated massless, the remaining quark-mass effects not explicitly accounted for in our formulae and treatments may be relatively power-suppressed in quark-mass ratios, but at least start only from two-loop order. These additional mass effects will not change the qualitative picture reported here, and we plan to have them analyzed quantitatively in detail in the future. 
}
upon setting $\mu=m_Q^{\mathrm{os}}$, there is no more explicit dependence on $m_Q^{\mathrm{os}}$ in the perturbatively-truncated ratio $\ZTA$ or $1-\ZTA = \mathrm{Z}_{\sigma}$, and the expression~\eqref{eq:TASmass_OSmass} reduces to just a function of $\alpha_s(m_Q^{\mathrm{os}})$ and $n_l$ (with the gauge group fixed).
Under this condition, the dependence of these ratios on the numerical value of $m_Q^{\mathrm{os}}$ enters solely via $\alpha_s(m_Q^{\mathrm{os}})$ in a perturbative series, which is thus essentially logarithmic.
The numbers for the electron were determined by inserting the benchmark input value $m_e^{\mathrm{os}} = 0.511$~MeV and the QED fine-structure constant $\alpha = 1/137$ into the Abelian counterpart of eq.~\eqref{eq:TASmass_OSmass}.
The inputs used for determining the numbers in the case of $t$-quark are the mass $m_t^{\mathrm{os}} = 172.69$~GeV\cite{ATLAS:2018fwq}\footnote{We set aside, for the moment, the dispute within the high-energy physics community over interpreting this value as the perturbative pole mass of $t$-quark.} and the perturbative QCD coupling value $\alpha_s^{(6)}(m_t^{\mathrm{os}}) = 0.1076$ $\MSbar$-renormalized in the 6-flavor scheme.
The conventional QCD scale uncertainty determined for the result $\SigmaMass^t = 159.0$~GeV for the $t$-quark is less than a permile.

The cases of $b$-quark and $c$-quark require extra care, as the numerical values for their perturbative pole masses are not determined very accurately at the moment (which are also not very stable against incorporation of high-order corrections in the perturbative treatments).
To extract the $\SigmaMass$ for them, which by itself is theoretically well-defined owing to the merits discussed in preceding paragraphs, we establish directly the perturbative relation between $\SigmaMass$ and $\mMS$ of a heavy quark up to three loops, which is provided in eq.~\eqref{eq:TASmass_MSmass}, where the argument of the logarithm is consistently rewritten in terms of the $\alpha_s$-renormalization scale $\mu$ and the particle's $\MSbar$ mass. 
This relation is thus very suitable for extracting the numerical value of $\SigmaMass$ directly from the better known $\mMS$ for the $b$-quark and $c$-quark, circumventing the explicit reference to their less-well-known $\mOS$.
More specifically, the external inputs used for determining the $\SigmaMass^{b}$ of the $b$-quark are the mass $\overline{m}_b(\mu=\overline{m}_b) = 4.18$~GeV~\cite{ParticleDataGroup:2024cfk} and the 5-flavor perturbative QCD coupling value $\alpha_s^{(5)}(\overline{m}_b) = 0.224$ (determined using a four-loop running from $\alpha_s^{(5)}(m_z) = 0.1179$).
The conventional QCD-scale uncertainty determined for the result $\SigmaMass^{b} = 3.96$~GeV of $b$-quark is $[-1.6\%,\,  +2.8\%]$.
In the case of the $c$-quark, the external inputs used for determining $\SigmaMass^{c}$ are the mass $\overline{m}_c(\mu=\overline{m}_c) = 1.27$~GeV~\cite{ParticleDataGroup:2024cfk} and the 4-flavor perturbative QCD-coupling value $\alpha_s^{(4)}(\overline{m}_c) = 0.38$.\footnote{This value is determined by using a three-loop $\alpha_s$-running from $\alpha_s^{(5)}(m_z) = 0.1179$ with the decoupling of $b$-quark around its mass scale, which was cross-checked against RunDec\cite{Chetyrkin:2000yt}.}
The QCD scale uncertainty for the result $\SigmaMass^{c} = 1.17$~GeV of $c$-quark is naively estimated to be about $[-5\%,\,  +5\%]$ by simply taking twice of its variation from the scale $\mu=\overline{m}_c$ to the scale $\mu=2\,\overline{m}_c$ (as the running of $\alpha_s$ is unstable or reliable below $1$~GeV).
Regarding the values of $\ZTA$ for $b$- and $c$-quark, we determine them by first rewriting $\mOS$ in the argument of the logarithm in the expression \eqref{eq:TASmass_OSmass} solely in terms of $\mMS$, i.e.~with the aid of essentially the inverse of eq.~\eqref{eq:OSmass_TASmass}.
Subsequently, this ratio $\ZTA$ 
can be computed without making any explicit reference to the numerical value of $\mOS$.
However, given the lack of precise values of $\mOS$ for $b$-quark and especially $c$-quark, one shall take a grain of salt when interpreting these numbers $\ZTA$ for them.

The most noticeable feature exhibited in the numbers in table~\ref{tab:tasmasses} is that the trace-anomaly contribution does make a considerable non-vanishing portion for the perturbative on-shell or pole mass of heavy quarks, and this portion grows quickly from about 8\% for $t$-quark to more than 30\% for $c$-quark.
In spite of the close analogue between the above discussion on the perturbative pole masses of heavy quarks and the non-perturbative proton state, there is, however, an important and interesting difference regarding the following aspect that should be kept in mind: 
the contribution from the trace-anomaly operator to the perturbative pole masses of heavy quarks would vanish in the limit of the vanishing Higgs-generated $\sigma$-mass term, essentially due to the chiral symmetry of the gauge interactions.   
Therefore, very formally speaking, in the limit of vanishing Higgs-generated fermion masses, the picture observed here interestingly resembles a bit that of the pion case~\cite{Yang:2014qna,Tanaka:2018nae,Liu:2023cse,Wang:2024lrm,Hoferichter:2025ubp} in the chiral (i.e.~all quarks massless) limit regarding this aspect.

\section{Conclusion}
\label{sec:conc}

The masses of elementary particles, such as the electron and quarks, are fundamental parameters of the Standard Model of particle physics, and therefore involved in the vast amount of theoretical predictions generated within its framework.  
A better understanding of the technical origin of mass for both fundamental and composite particles --- particularly the proton, which remains one of the most important and intriguing quantities in particle physics --- is of utmost importance; 
such understanding is critical not only for deepening our theoretical understanding in the quantum dynamics of the Standard Model but also for enabling its high-precision tests.
Due to the intimate connection between the total EMT and the rest mass for an isolated system, as anticipated on general physical grounds, studying the properties of the EMT within the context of QFT offers a particularly interesting and fruitful perspective on this matter, especially in view of the intriguing quantum anomaly discovered in its trace.

In this work we have presented a novel direct diagrammatic proof of the \textit{identity} between the forward matrix element of the EMT-trace operator over a single particle's on-shell state, and its perturbative pole mass --- \textit{defined} by the zero-point of the inverse propagating function ---
to any loops in perturbative gauge theories with gauge bosons and fermions, and to all orders in the dimensional regulator, 
without appealing to any pre-laid operator renormalization conditions or Ward identities. 
Our proof is based on the equation of mass-dimensional analysis in dimensional regularization, the topological properties of contributing Feynman diagrams, and the on-shell renormalization condition. 
It is thus helpful to consolidate the theoretical basis of the aforementioned identity in a transparent manner, with a particular emphasis on the role played by the trace-anomaly operator.

After proving the identity exactly for elementary fermions to any loops in perturbative gauge theories, we performed explicit perturbative calculations, up to three loops, to show that there are indeed non-vanishing contributions from the trace-anomaly operator to the perturbative pole masses of the on-shell electron and heavy $t$-,\,$b$-,\,$c$-quarks.
On top of this, observing interestingly that the trace-anomaly contribution contains all leading-IR-renormalon effects up to three loops, we propose accordingly a new scheme- and scale-independent trace-anomaly-subtracted $\sigma$-mass definition for these elementary particles.

To gain a more concrete understanding of this mass, we have taken a preliminary look at the numerical values of the so-defined $\sigma$-mass for the $t$-,\,$b$-,\,$c$-quarks, observing several intriguing features.   
Therefore, beyond the immediate utility of these findings in applications to high-energy physics involving heavy quarks (to be explored in future work), we hope that comparing the behaviors of EMT matrix elements --- between elementary and composite particles, and between perturbative and non-perturbative treatments --- may yield deeper insights into the properties of the EMT and the origin of masses for quantum objects.   

\section*{Acknowledgements}

The work of L.~C. was supported by the Natural Science Foundation of China under contract No.~12205171, No.~12235008, No.~12321005, and grants from Department of Science and Technology of Shandong province tsqn202312052 and 2024HWYQ-005.
The authors gratefully acknowledge the valuable discussions and insights provided by the members of the China Collaboration of Precision Testing and New Physics.

\appendix

\section{Three-loop transformations relating the $\SigmaMass$ mass to the perturbative pole mass and $\MSbar$ mass for heavy quarks}
\label{ApxSec:addmassrelations}

The explicit three-loop perturbative result for the ratio $\SigmaMass/\mOS$ is given in eq.~\eqref{eq:TASmass_OSmass},  
which may be used to obtain the numerical values for $\SigmaMass$ from the given perturbative pole mass $\mOS$, provided the latter is known to sufficient accuracy for the fermion in question.
The perturbative inverse of this relation is derived below in eq.~\eqref{eq:OSmass_TASmass}, where the mass dependence in the logarithm is consistently rewritten in terms of $\SigmaMass$ as in $\logusm \equiv \ln\big(\mu^2/\SigmaMass^2 \big)$.
This result can thus be employed to conveniently transform the original perturbative expression for a physical observable involving heavy quarks with the perturbative pole mass $\mOS$ into a function of the heavy quarks' $\SigmaMass$.
{\fontsize{9pt}{15}\selectfont
\begin{align} \label{eq:OSmass_TASmass}
\frac{\mOS}{\SigmaMass} = {}
&
1 
+ \api\,\frac{3}{2}\, C_F 
+ \Bigg(\api \Bigg)^2 \frac{1}{48}\, 
\Big[ 
   -26\,C_F + 185\,C_A C_F + 9\,C_F^2 - 26\,C_F n_l \nn
&
- 12\,C_F \logusm 
   + 66\,C_A C_F \logusm 
   - 12\,C_F n_l \logusm 
\Big] \nn
&
+ \Bigg(\api \Bigg)^3 \frac{1}{1728}\, 
\Big[
   788\,C_F - 9328\,C_A C_F + 26486\,C_A^2 C_F - 891\,C_F^2 \nn
&
- 6048\,C_A C_F^2 + 3483\,C_F^3 + 1144\,C_F n_l 
   - 6952\,C_A C_F n_l - 891\,C_F^2 n_l \nn
&
+ 356\,C_F n_l^2 - 96\,C_F \pi^2 + 624\,C_A C_F \pi^2 
   - 528\,C_A^2 C_F \pi^2 - 360\,C_F^2 \pi^2 \nn
&
+ 1980\,C_A C_F^2 \pi^2 - 48\,C_F n_l \pi^2 
   - 168\,C_A C_F n_l \pi^2 - 360\,C_F^2 n_l \pi^2 \nn
&
+ 48\,C_F n_l^2 \pi^2 + 576\,C_F^2 \pi^2 \log(2) 
   + 576\,C_F^2 n_l \pi^2 \log(2) \nn
&
+ 1584\,C_A^2 C_F \pi^2 \log(2) 
   - 288\,C_A C_F \pi^2 \log(2) 
   - 396\,C_A C_F^2 \pi^2 8\log(2) \nn
&
- 24\,C_A C_F n_l \pi^2 \log 4096 
   + 312\,C_F \logusm - 4476\,C_A C_F \logusm \nn
&
+ 14046\,C_A^2 C_F \logusm - 432\,C_F^2 \logusm 
   + 594\,C_A C_F^2 \logusm \nn
&
+ 624\,C_F n_l \logusm - 4476\,C_A C_F n_l \logusm 
   - 432\,C_F^2 n_l \logusm \nn
&
+ 312\,C_F n_l^2 \logusm + 72\,C_F \logusm^2 
   - 792\,C_A C_F \logusm^2 \nn
&
+ 2178\,C_A^2 C_F \logusm^2 
   + 144\,C_F n_l \logusm^2 - 792\,C_A C_F n_l \logusm^2 \nn
&
+ 72\,C_F n_l^2 \logusm^2 
   - 864\,C_A C_F \zeta_3 - 2376\,C_A^2 C_F \zeta_3 \nn
&
+ 432\,C_F^2 \zeta_3 + 4752\,C_A C_F^2 \zeta_3 
   - 864\,C_A C_F n_l \zeta_3 \nn
&
+ 432\,C_F^2 n_l \zeta_3 
\Bigg]
\end{align}
}
In addition, we provide in eq.~\eqref{eq:TASmass_MSmass} the explicit three-loop perturbative result for the ratio $\SigmaMass/\mMS$, where $\logums \equiv \ln\big(\mu^2/\mMS^2 \big)$ denotes the logarithm of the squared ratio of the $\alpha_s$-renormalization scale $\mu$ over the heavy quark's $\MSbar$ mass $\mMS$. 
(The transformation relation between the on-shell and $\MSbar$ mass is needed up to three loops, which is taken from refs.~\cite{Melnikov:2000qh,Chetyrkin:1999qi}, and see refs.~\cite{Marquard:2015qpa,Marquard:2016dcn,Kataev:2019zfx} for higher-order results.)
This relation is thus suitable to obtain the numerical value of $\SigmaMass$ directly from the known $\mMS$ for the given heavy quark, circumventing the explicit reference to the $\mOS$ which may be particularly desirable in case its numerical value is not determined with sufficient accuracy due to the IR-renormalon issue in the fixed-order perturbative relations.
{\fontsize{9pt}{15}\selectfont
\begin{align} \label{eq:TASmass_MSmass}
\frac{\SigmaMass}{\mMS} = {}
&
1 
+ \api\, \frac{1}{4}\,C_F \Big[-2 + 3 \logums\Big]
\Bigg\{
6 C_F \Big(-4 C_F - 3 C_F \logums\Big) 
\nn
& 
+ \Bigg(\api \Bigg)^2 \frac{1}{16}\, 
 + \frac{1}{24} C_F \,
\Bigg[
 -78 + 579 C_F 
 + 66 n_l  + 32 \pi^2 + 120 C_F \pi^2 
\nn
&
 - 16 n_l \pi^2 
- C_A \pi^2 \big(32 - 96 \log(2)\big) 
 - 192 C_F \pi^2 \log(2) 
 - \Big(-212 C_A 
 \nn
 &
 + 108 C_F + 8(1+n_l)\Big) \logums 
 + 12 \Big(11 C_A + 9 C_F - 2(1+n_l)\Big) \logums^2 
 \nn
&
 + 288 C_F \zeta_3 
 - 9 C_A \Big(41 + 16 \zeta_3\Big)
\Bigg]
\Bigg\}
+ \Bigg(\api \Bigg)^3 \frac{1}{64}\, 
\Bigg\{
 -36 C_F^2 \big(-4 C_F 
 \nn
&
 - 3 C_F \logums\big) 
+ \frac{1}{24} C_F\,
\Big[
 -2 \big(-212 C_A 
 + 108 C_F + 8(1+n_l)\big)\big(-4 C_F 
 \nn
& 
 - 3 C_F \logums\big)
 + 48 \big(11 C_A + 9 C_F - 2(1+n_l)\big)\logums \big(-4 C_F - 3 C_F \logums\big)
\Big]
\nn
& 
 + 3 C_F \Big(
 -C_F \big(4 + 3 \logums\big)\big(4 C_F + 3 C_F \logums\big) 
 + \frac{1}{12} C_F 
   \Big(
  286 + 21 C_F 
\nn
&
 + 142 n_l - 32 \pi^2 - 120 C_F \pi^2 + 16 n_l \pi^2 
 + 192 C_F \pi^2 \log(2) 
 + 4 \big( 63 C_F
 \nn
&
-185 C_A
 + 26(1+n_l)\big)\logums
 - 12 \big(11 C_A - 9 C_F - 2(1+n_l)\big)\logums^2         
\nn
&
- 288 C_F \zeta_3  
+ C_A \Big(-1111 + \pi^2(32 - 96 \log(2)) + 144 \zeta_3\Big)
 \Big)
 \Big)
 \nn
 &
+ \frac{1}{9720} C_F \,
\Bigg[
 540 \big(889 C_A^2 
+ 378 C_A C_F - 81 C_F^2 - 218 C_A(1+n_l)
\nn
&
  - 54 C_F(1+n_l) + 4(1+n_l)^2\big) \logums^2
 + 540 \big(242 C_A^2 
 + 297 C_A C_F 
 + 81 C_F^2 
 \nn
&
 - 88 C_A(1+n_l) 
- 54 C_F(1+n_l) + 8(1+n_l)^2\big) \logums^3
  + 45 \logums 
\Big(
\nn
&
 -16 (1+n_l)\big(-41 + 24 \pi^2 + n_l(67 - 12 \pi^2)\big) 
+ 88 C_A^2 \big(-73 
\nn
& 
+ 24 \pi^2(-1+3\log(2)) - 108 \zeta_3\big) 
 - 81 C_F^2 \big(-357 + 8 \pi^2(-5
 \nn
&   
 +8\log(2)) - 96 \zeta_3\big)      
+ 18 C_F \Big(-691 + 32 \pi^2(-1+4\log(2)) + 96 \zeta_3 
\nn
&
 - C_A \Big(63 C_F \big(-227 + 16 \pi^2(-7+10\log(2)) - 240 \zeta_3\big)
 + n_l(-475 
 \nn
& 
 + 8 \pi^2(-13+16\log(2)) + 96 \zeta_3)\Big)
 + 32 \big(47 + 6 \pi^2(-13+6\log(2)) 
\nn
&
+ 108 \zeta_3 
 + n_l(-250 + 3 \pi^2(7+\log 4096) + 108 \zeta_3)\big)\Big)
 \Big)
\nn
& 
+ 4\Big[
  -23515 - 43790 n_l - 20275 n_l^2 + 7488 \pi^2 - 360 n_l \pi^2 
\nn
&    
      + 360 n_l^2 \pi^2 
      - 23760 \zeta_3 - 8640 n_l \zeta_3 + 15120 n_l^2 \zeta_3 
\nn
&
      - 18 C_F \Big(182 \pi^4 - 40 \pi^2(139 - 168 \log(2) + 12 \log^2 2) 
\nn
&      
      + 11520 \operatorname{Li}_4\big(\tfrac{1}{2}\big)
      + 15 (643 + 32 \log^4 2 - 492 \zeta_3) 
\nn
& 
      + n_l(-238 \pi^4 + 60 \pi^2(-5+4\log(2))^2 + 11520 \operatorname{Li}_4\big(\tfrac{1}{2}\big) 
\nn
& 
      + 15(643 + 32 \log^4 2 + 804 \zeta_3))\Big) 
      + 810 C_F^2\big(4 \pi^4 + 2304 \operatorname{Li}_4\big(\tfrac{1}{2}\big)
\nn
&      
       + \pi^2(643 - 1440 \log(2) - 96 \log^2 2 + 12 \zeta_3) 
\nn
& 
      + 2(55 + 48 \log^4 2 + 522 \zeta_3 - 60 \zeta(5))\big)
      \Big]
\nn
&
      + 5 C_A^2 \Big[
      -584447 + 6444 \pi^4 - 19008 \log^4 2 - 456192 \operatorname{Li}_4\big(\tfrac{1}{2}\big) 
\nn
& 
      - 409104 \zeta_3 - 36 \pi^2(-3011 - 5520 \log(2) + 1056 \log^2 2 
\nn
&
      + 3168 \log(2) + 2754 \zeta_3) + 252720 \zeta(5)
      \Big]
\nn
& 
      + C_A \Big[
      4 n_l \big(-684 \pi^4 + 180 \pi^2(-91 - 264 \log(2) + 48 \log^2 2 
\nn
&       
      + 12 \log 4096) + 103680 \operatorname{Li}_4\big(\tfrac{1}{2}\big) 
      + 5(54373 + 864 \log^4 2 
\nn
& 
      + 5940 \zeta_3)\big) - 45 C_F \big(-78351 + 2080 \pi^4 - 768 \log^4 2 
\nn
& 
      - 18432 \operatorname{Li}_4\big(\tfrac{1}{2}\big) 
      - 70704 \zeta_3 - 24 \pi^2(-1193 + 320 \log(2) 
\nn
&      
      + 496 \log^2 2 + 1056 \log(2) + 684 \zeta_3) + 38880 \zeta(5)\big) 
\nn
& 
      + 4 \Big(3096 \pi^4 + 103680 \operatorname{Li}_4\big(\tfrac{1}{2}\big) - 540 \pi^2(-345 + 512 \log(2) 
\nn
& 
      + 8 \log^2 2 - 48 \log(2) + 18 \zeta_3) 
      + 5(22945 + 864 \log^4 2 
\nn
&
      + 27324 \zeta_3 + 9720 \zeta(5))\Big)
      \Big]
   \Bigg]
\Bigg\}
\nn
\end{align}
}

\section{Perturbative results for $Z_3^{os}$ and $Z_{\psi,q}^{os}$ up to three loops}
\label{ApxSec:Z3Z3res}

Along the way of performing the aforementioned explicit checks up to three loops, we derived, as a bonus of this computation, the three-loop QCD result for the on-shell wave-function renormalization constant $Z_3^{os}$ for the gluon and $Z_{\psi,q}^{os}$ for massless quarks, the final missing ingredients for a complete three-loop on-shell renormalization in QCD with one massive quark.\footnote{We thank Florian Herren for informing us, albeit only after our derivation of the explicit expressions for $Z_3^{os}$ and $Z_{\psi,q}^{os}$ had been completed, that these expressions were involved in the computations 
in refs.~\cite{Chetyrkin:1997un,Gerlach:2018hen}.}
We have kept the $\epsilon$-dependence in perturbative coefficients to high orders sufficient for three-loop QCD calculations.
We observe that the three-loop on-shell gluon wave-function renormalization constant $Z_3^{os}$, given in eq.~\eqref{eq:z3os3L}, starts to exhibit dependence on the gauge-fixing parameter $\xi$ from three loops. 
The $\xi$-dependence in the on-shell massive quark wave-function renormalization constant with dimensional regularization was observed in ref.~\cite{Melnikov:2000zc}, which only starts to appear from three loops.
In eq.~\eqref{eq:z2os3L}, we provide the counterpart for a massless quark field, where $\xi$-dependence also appears from three loops.
We note in passing that the coefficient of the highest pole $\epsilon^{-3}$ in this perturbative expression is proportional to $\xi-1$, and hence vanishes in the Landau gauge, indicating its IR-origin.
The results presented in eq.~\eqref{eq:z3os3L} and eq.~\eqref{eq:z2os3L} were derived using two independent set-ups with different techniques, one directly working at the on-shell kinematic point evaluating analytically only vacuum integrals~\cite{Schroder:2005va} and the other approaching the on-shell limit from generic off-shell kinematics (where the AMFlow method~\cite{Liu:2017jxz,Liu:2020kpc,Liu:2021wks,Liu:2022mfb,Liu:2022chg} is employed to evaluate the integrals).
As expected, in the end, a perfect agreement was found between the two computational approaches. 
{
\fontsize{9pt}{15}\selectfont
\begin{align} \label{eq:z3os3L}
Z_{3}^{os} ={}
&
1 +  
\api 
\Bigg\{ 
-\frac{1}{ \epsilon} \frac{T_F}{3} 
- \frac{1}{3} T_F \loguos 
-\epsilon  \frac{T_F}{36}   
\Big[ 
\pi^2 + 6 \loguos^2 
\Big] 
+ \epsilon^2\frac{T_F}{36}  
\Big[ 
-\pi^2 \loguos 
\nn
&
- 2 \loguos^3 + 4 \zeta_3 
\Big]
\Bigg\} 
+ \Bigg(\api \Bigg)^2 
\Bigg\{
\frac{1}{ \epsilon^2}\frac{T_F}{144}
\Big[
35\, C_A - 16\, n_l T_F
\Big] 
+ \frac{1}{\epsilon} \frac{T_F}{288}  
\Big[
\nn
&
-9 (5 C_A + 4 C_F) 
+ 52 C_A \loguos 
- 32 (-1 + n_l) T_F \loguos 
\Big] 
+ \frac{T_F }{1728} \,
\Big[
\nn
&
+ 13 C_A (9 + 2 \pi^2) 
- 4 \big( 405 C_F + 4 (-1 + n_l) \pi^2 T_F \big) 
- 108 (5 C_A 
\nn
&
+ 4 C_F) \loguos
+ 48 \big( C_A 
- 2 (-3 + n_l) T_F \big) \loguos^2 
\Big] 
+ \epsilon \frac{ T_F  }{3456} 
\Big[
-36 C_F (93 
+ 2 \pi^2) 
\nn
&
+ 4 \big( -1620 C_F 
+ C_A (117 + 4 \pi^2)
- 8 (-3 + n_l) \pi^2 T_F \big) \loguos 
- 216 (5 C_A 
\nn
&
+ 4 C_F) \loguos^2 
- 16 \big( 7 C_A + 4 (-7 + n_l) T_F \big) \loguos^3 
+ 128 (-1 + n_l) T_F \zeta_3 
\nn
&
- C_A \big( 507 
+ 90 \pi^2 + 208 \zeta_3 \big) 
\Big] 
\Bigg\} 
\nn
&  
+ \Bigg(\api \Bigg)^3
\Bigg\{
-\frac{1 }{\epsilon^3}\frac{T_F }{3456 }
\Big[ 
-8 C_A (3 + 76 n_l) T_F + 128 n_l^2 T_F^2 + C_A^2 (695 + 9 \xi) \Big]  
\nn
& 
+ \frac{1}{\epsilon^2}\frac{T_F}{6912}
\Big[ 
-96 C_F (1 + 8 n_l) T_F + 40 C_A (21 C_F - 5 T_F - 28 n_l T_F) 
\nn
& 
+ C_A^2 (2015 + 63 \xi)  
- 2 \big( -8 C_A (-61 + 70 n_l) T_F + 128 (-2 + n_l) n_l T_F^2 
\nn
&
+ C_A^2 (545 + 27 \xi) \big) \loguos 
\Big]
- \frac{1}{ \epsilon} \frac{T_F}{41472 } 
\Big[
-16 \Big( 
27 C_F^2 + C_F (876 - 678 n_l) T_F 
\nn
&
- 8 (-2 + n_l) n_l \pi^2 T_F^2 
\Big)   
- 18 \Big( 
96 C_F (3 - 4 n_l) T_F + 8 C_A (39 C_F + 45 T_F 
\nn
&  
- 70 n_l T_F) 
+ C_A^2 (811 + 63 \xi) 
\Big) \loguos  
+ 6 \Big( 
-56 C_A (-21 + 10 n_l) T_F
\nn
& 
+ 128 (2
 - 6 n_l + n_l^2) T_F^2
+ C_A^2 (491 + 81 \xi) 
\Big) \loguos^2  
+ C_A^2 \Big( 
\pi^2 (545 + 27 \xi) 
\nn
&
+ 3 \big( 4897 + 411 \xi - 1296 \zeta_3 \big) 
\Big) 
+ 8 C_A \Big(
- T_F\big( 111 - 61 \pi^2 + 70 n_l (12 
\nn
& 
+ \pi^2) \big) 
+ 3 C_F \big(-875 + 216 \zeta_3 \big) 
\Big) 
\Big]  
+ \frac{T_F}{1244160} 
\Big[ 
270 \Big( 
288 C_F (5 - 2 n_l) T_F 
\nn
& 
- 40 C_A (3 C_F - 47 T_F + 22 n_l T_F) 
+ C_A^2 (569 + 189 \xi) 
\Big) \loguos^2  
- 60 \Big(
8 C_A (
\nn
&
+193 
- 106 n_l) T_F
 + 128 (12 - 14 n_l + n_l^2) T_F^2 
+ C_A^2 (1121 + 243 \xi) 
\Big) \loguos^3  
\nn
& 
+ 240 \Big( 
C_F T_F \big(7234 - 522 \pi^2 + n_l (-430 + 216 \pi^2) 
 - 567 \zeta_3 \big) 
+ 9 C_F^2 \big(
 \nn
& 
-77 + 24 \pi^2 (-5 + 8\log(2)) - 3 \zeta_3 \big) 
+ 64 (-2 + n_l) n_l T_F^2 \zeta_3 
\Big) 
+ C_A^2 \Big(
\nn
& 
+548615 - 11448 \pi^4 + 51840 \log ^4(2)
+ 1244160 \mathrm{Li}_4\big( \tfrac12 \big) + 552750 \zeta_3 
\nn
& 
+ 45 \pi^2 (811 + 63 \xi - 1152 \log^2(2)) 
- 315 \xi (-511 + 72 \zeta_3) 
\Big) 
- 30 \loguos \Big( 
\nn
& 
+16 \big(
-81 C_F^2 + 18 C_F (-236 + 23 n_l) T_F 
+ 8 (2 - 6 n_l + n_l^2) \pi^2 T_F^2
\big) 
\nn
& 
+ C_A^2 \big( 
\pi^2 (491 + 81 \xi) 
+ 9 (4325 + 411 \xi - 1296 \zeta_3) 
\big)  
+ 8 C_A \big(
- (99
\nn
& 
 + 2286 n_l - 147 \pi^2 + 70 n_l \pi^2) T_F 
+ 9 C_F ( 115 + 216 \zeta_3) 
\big) 
\Big) 
+ 4 C_A \Big( 
\nn
& 
-10 T_F \big(
-6361 - 405 \pi^2 
+ 2 n_l (4724 + 315 \pi^2 - 888 \zeta_3) 
+ 6285 \zeta_3 
\big) 
\nn
& 
+ 3 C_F \big( 
1584 \pi^4 + 90 \pi^2 (77 - 192 \log(2) + 96 \log^2(2)) \nn
& 
- 5 (5096 + 1728 \log^4(2) + 41472 \mathrm{Li}_4\big(\tfrac12\big) 
+ 47655 \zeta_3)
\big) 
\Big) 
\Big]
\Bigg\} 
\end{align}
}

{
\fontsize{9pt}{15}\selectfont
\begin{align} \label{eq:z2os3L}
Z_{\psi,q}^{os} = {}&
 1 
+ \Bigg(\api\Bigg)^2 
\Bigg\{
\frac{1}{\epsilon} \frac{C_F T_F}{16} 
- \frac{C_F T_F}{96} \Big[ 5 - 12 \loguos \Big] 
+ \epsilon \frac{C_F T_F}{576} 
\Big[ 89 + 6 \pi^2  
\nn
&
- 60 \loguos
+ 72 \loguos^2 \Big]
\Bigg\} 
+ \Bigg(\api\Bigg)^3 
\Bigg\{
-\frac{1}{\epsilon^3} 
\frac{C_A C_F T_F (-1 + \xi)}{192} 
+ \frac{1}{\epsilon^2} \frac{1}{576} 
\Big[
\nn
& 
-56 C_A C_F T_F 
+6 C_F^2 T_F
 + 8 C_F T_F^2 
+ 16 C_F n_l T_F^2 
+ 9 C_A C_F T_F \xi  
\nn
&
+ 9 C_A C_F T_F \loguos
- 9 C_A C_F T_F \xi \loguos
\Big] 
+ \frac{1}{\epsilon} 
 \frac{1}{6912} 
\Big[
- 108 C_F^2 T_F 
\nn
&
+1568 C_A C_F T_F 
+ 9 C_A C_F \pi^2 T_F 
- 80 C_F T_F^2 - 160 C_F n_l T_F^2 
\nn
&
- 420 C_A C_F T_F \xi - 9 C_A C_F \pi^2 T_F \xi 
-( 1224 C_A C_F T_F  - 216 C_F^2 T_F  
\nn
&
- 288 C_F n_l T_F^2 
- 324 C_A C_F T_F \xi) \loguos 
+ 162C_A C_F T_F ( 1 -  \xi) \loguos^2
\Big] 
\nn
&
+ \frac{1}{20736} 
\Big[
-9632 C_A C_F T_F + 7974 C_F^2 T_F - 306 C_A C_F \pi^2 T_F
\nn
&
+ 54 C_F^2 \pi^2 T_F 
- 280 C_F T_F^2 
- 560 C_F n_l T_F^2 + 72 C_F n_l \pi^2 T_F^2 
\nn
& 
+ 4884 C_A C_F T_F \xi 
+ 81 C_A C_F \pi^2 T_F \xi 
+ (
12132 C_A C_F T_F  
\nn
&
 +81 C_A C_F \pi^2 T_F 
- 972 C_F^2 T_F    
- 720 C_F n_l T_F^2 
- 3780 C_A C_F T_F \xi  
\nn
&
- 81 C_A C_F \pi^2 T_F \xi)\loguos 
- (3132 C_A C_F T_F  
+ 972 C_F^2 T_F 
- 864 C_F T_F^2  
\nn
& 
+ 432 C_F n_l T_F^2 
- 3132 C_A C_F T_F) \loguos^2 
+ 486 C_A C_F T_F (1-\xi) \loguos^3 
\nn
& 
+ 54 C_A C_F T_F \psi^{(2)}(1) 
- 54 C_A C_F T_F \xi \,\psi^{(2)}(1) 
- 1728 C_A C_F T_F \zeta_3
\nn
&
 + 2592 C_F^2 T_F \zeta_3 
- 864 C_A C_F T_F \xi \,\zeta_3
\Big]
\Bigg\}
\end{align}
}

\bibliographystyle{utphysM}
\balance
\biboptions{sort&compress}

\bibliography{EMTtraceMass} 


\end{document}